\documentclass[dvips]{article}
\usepackage{icrctc07}
\title{X-Ray Observations of LSI+61303 with Swift}
\shorttitle{Observations of LSI+61303 with Swift}
\authors{J. Holder$^{1}$, A. Falcone$^{2}$ and D. Morris$^{2}$.}
\shortauthors{J. Holder, A. Falcone and D. Morris}
\afiliations{$^1$Bartol Research Institute, Department of Physics and Astronomy, University of Delaware, Newark, DE 19716, USA\\ $^2$Department of Astronomy and Astrophysics, Pennsylvania State University, University Park, PA 16802,  USA}
\email{jholder@physics.udel.edu}

\abstract{The TeV emitting high-mass X-ray binary system LSI+61303 was observed with the Swift satellite from early September 2006 to early January 2007. Many of these observations were contemporaneous with TeV observations. The data consist of observations on 24 separate days with durations ranging between 700s and 4700s, and partially cover 4.5 orbital periods of the binary system. We present here an analysis of the 0.2 to 10keV X-ray data from the Swift-XRT instrument. Contemporaneous optical data from UVOT are also available.}

\begin{document}
\maketitle

\section{Introduction}
LS I +61 303 is a high mass X-ray binary system at a distance of $\sim
2~\mathrm{kpc}$, composed of a B0 Ve star with a circumstellar disc
and a compact object. The observed radio through optical emission is
modulated with a period ($P=26.4960\pm0.0028~\mathrm{days}$) believed
to be associated with the orbital period of the binary system
\cite{gregory02, casares05}. Periastron takes place at phase 0.23 and
the eccentricity is $0.72\pm0.15$ (although see \cite{grundstrom07}
for a recent re-evaluation of the orbital elements). 

The detection of an extended and apparently precessing jet-like
radio-emitting structure led to the possible identification of LS I
+61 303 as a microquasar, with the emission generated through particle
acceleration in a relativistic accretion-driven jet
\cite{massi04}. More detailed AU-scale observations \cite{dhawan07}
strongly support a counter-model, where the radio emission arises from
particles shock-accelerated in the interaction of a pulsar wind with
the circumstellar material. The absence of any features in the X-ray
spectrum (e.g. thermal components due to an accretion disc) also
support the pulsar wind model. However, Romero et al. \cite{Romero07}
argue that the relative wind strengths are such that it is not
possible to produce the simple elongated shape observed in the VLBA
images, and that the gamma-ray lightcurve is more easily explained as
the result of variable accretion onto a compact object.

At higher energies, LS I +61 303 was associated with the COS-B source
2CG 135+01 \cite{swanenburg81}, and the EGRET source 2EG J0241+6119
\cite{kniffen97}.  More recently, LS I +61 303 has been discovered to
be a strong source (peak flux $\sim10\%$ of the steady flux from the
Crab Nebula) of very high energy gamma-rays
($>100~\mathrm{GeV}$)\cite{albert06, maier07}; one of only three known
galactic sources of variable TeV emission. The particle acceleration
and photon generation mechanisms which produce the TeV emission are
not yet clear, and detailed contemporaneous multiwavelength
observations will likely be required to resolve the situation.

Previous X-ray observations of this source have been made with ROSAT,
ASCA, BeppoSAX, RXTE, XMM, INTEGRAL and Chandra (see
\cite{chernyakova06} for a summary), revealing complex behaviour with
significant variation of flux and photon spectral index on both short
($\sim1000~\mathrm{s}$) and longer (monthly) timescales. The detection
of a TeV signal from LS I +61 303 by the VERITAS telescope array in
fall 2006 led us to propose target of opportunity observations with
the Swift satellite. The resulting dataset, combined with some earlier
Swift observations, provides the most well-sampled long-term
X-ray/optical/TeV monitoring which has been performed on this
source. In this paper we focus on some initial results obtained with
the Swift X-ray Telescope. A complementary paper discusses the
relationship between these observations and the TeV emission
\cite{smith07}. We also note that an analysis reaching similar results
and conclusions has recently been presented by Esposito et al
\cite{esposito07}.

\begin{figure*}[]
\begin{center}
\includegraphics [width=0.52\textwidth]{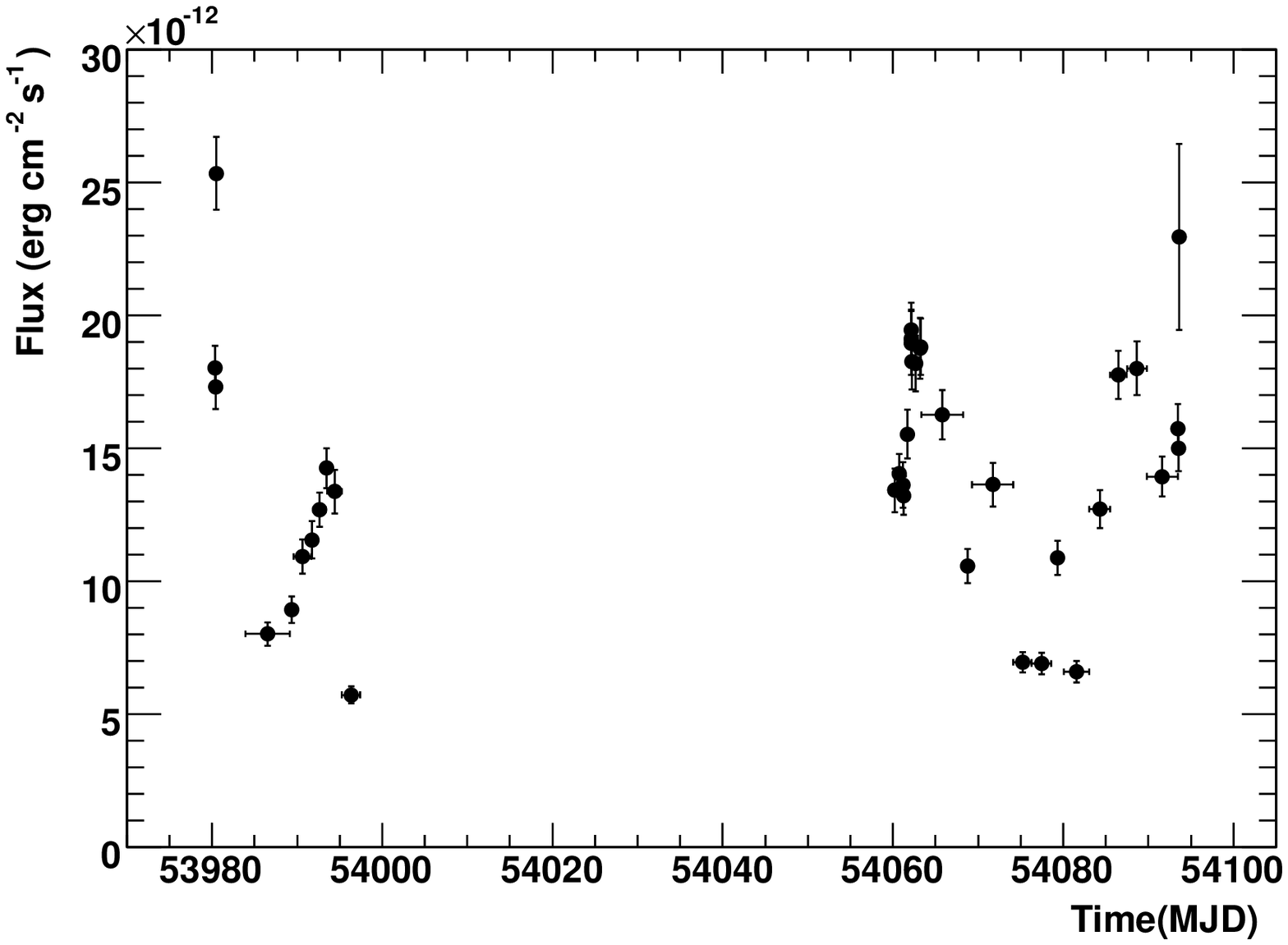}\includegraphics [width=0.52\textwidth]{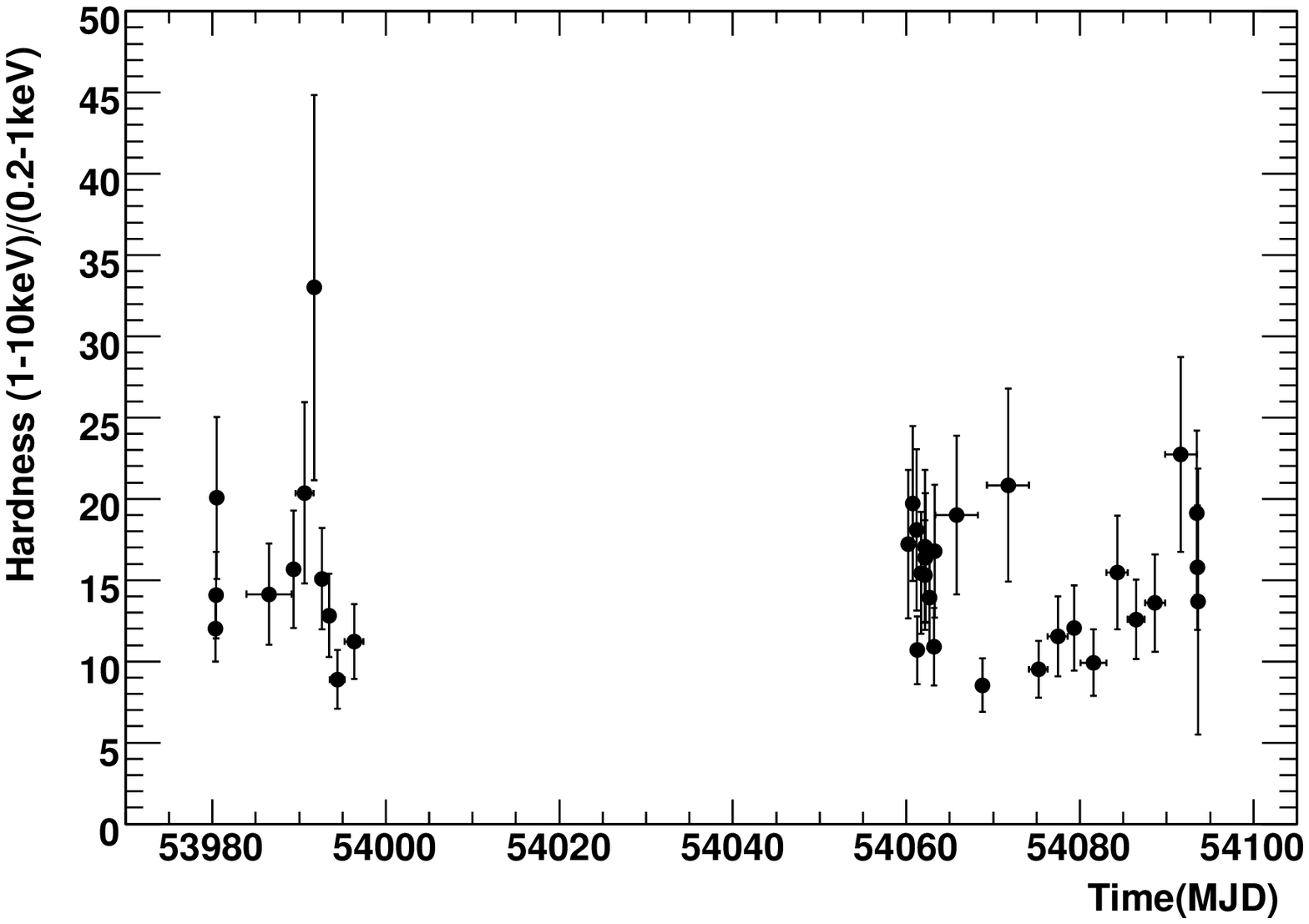}
\end{center}
\caption{Lightcurve for Swift observations of LS I +61 303. The left panel shows the XRT count rate (0.2-10~keV) as a function of modified Julian date, with an average flux of $1.4\times10^{-11}\mathrm{erg~cm^{-2}~s^{-1}}$. The right panel shows the ratio of the 1-10 keV flux over the 0.2-1 keV flux.}\label{lightcurve}
\end{figure*}

\section{Observations and Data Analysis}
The Swift satellite carries three instruments: the Burst Alert
Telescope (BAT) operates between 15-150~keV and is used primarily as a
gamma-ray burst trigger; the X-ray Telescope (XRT) is an imaging
instrument operating from 0.2-10~keV with a single photon point spread function of 18~arcsec (half-power diameter) and a spectral resolution of 140~eV at 5.9~keV (at launch);
UVOT, the UV/Optical telescope carries optical filters to cover the
range 170-650~nm. The primary mission of the satellite is to
investigate gamma-ray bursts and their afterglows; however, between
gamma-ray burst observations Swift is an excellent tool which can provide
observations of other astrophysical X-ray sources,
particularly when fast response and/or frequent slews are required.

The XRT observations were made in photon counting mode on 24 separate
days with exposures ranging between 700s and 4700s, covering a period
from early September 2006 to late December 2006 and corresponding to
a total exposure of 56~ksecs. The period following the ToO trigger
(from November 21st to December 24th) is particularly well-sampled,
with observations approximately once every two days.

The XRT data were processed using the most recently available standard
Swift tools: xrt software version 2.0.1, ftools version 6.2, and XSPEC
version 12.3.1.  The ancillary response files were generated with the
xrtpipeline task xrtmkarf. The spectral analysis presented in this
paper used XRT data extracted from the 0.3-10 keV energy band, thus
avoiding instrument response uncertainties present below 0.3 keV.

\section{Results and Discussion}
\begin{figure*}[]
\begin{center}
\includegraphics [width=0.52\textwidth]{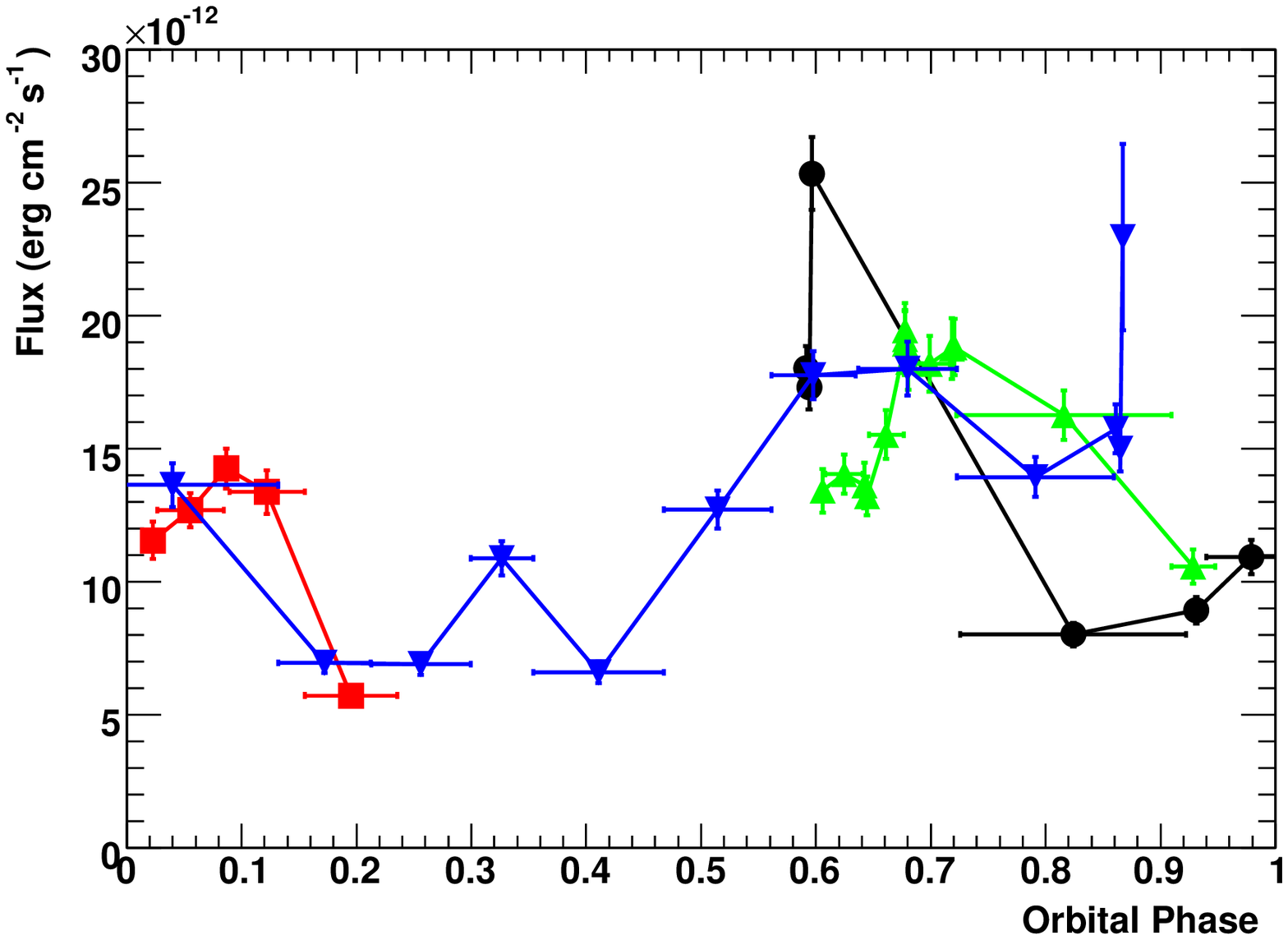}\includegraphics [width=0.52\textwidth]{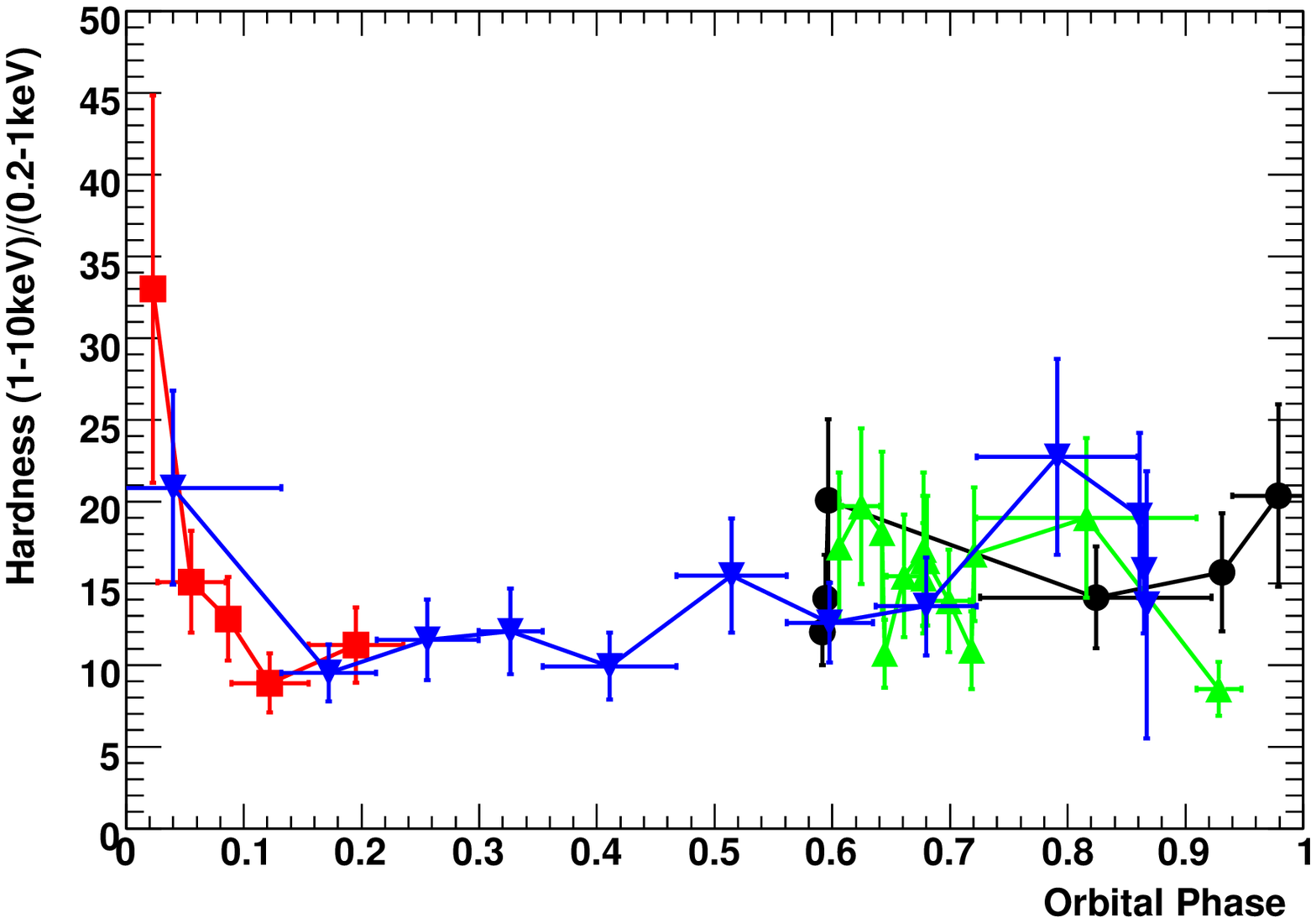}
\end{center}
\caption{Lightcurve for Swift observations of LS I +61 303. The left panel shows the XRT count rate (0.2-10keV) as a function of orbital phase, the right panel shows the ratio of the 1-10 keV flux over the 0.2-1 keV flux, also as a function of orbital phase}\label{phase}
\end{figure*}

Figure~\ref{lightcurve} shows the lightcurve and hardness ratio as a
function of time for the XRT observations. The XRT count rate to flux
conversion factor used, obtained using the mean spectral fit
parameters for these data, was
$5.7\times10^{-11}\mathrm{erg~cm^{-2}~s^{-1}/counts~s^{-1}}$. The
systematic error associated with any spectral variations is not
included in the error bars.  Figure~\ref{phase} shows the lightcurve
and hardness ratio as a function of orbital phase for these same
observations, using the orbital elements from \cite{casares05}. Points
indicated by the same symbol and colour correspond to measurements
taken during the same orbital cycle. Note that the lines between
points are drawn only to help illustrate this - the short timescale
variability is such that this simple interpolation can not be used to
estimate the emission level between observations.

The lightcurve shows clear variability in the flux of a factor of 4 -
5. While there are elevated states around orbital phases 0.6 - 0.8 and
0.0 - 0.1, it is evident that the emission level does not follow
precisely the same pattern each orbital cycle. The variation in the
hardness ratio is not statistically significant:
$\chi^2$/d.o.f.=45/36 for a constant fit, corresponding to a chance
probability of 13\%.

\begin{figure*}[]
  \begin{center}
    \includegraphics [height=6.8cm, angle=-90]{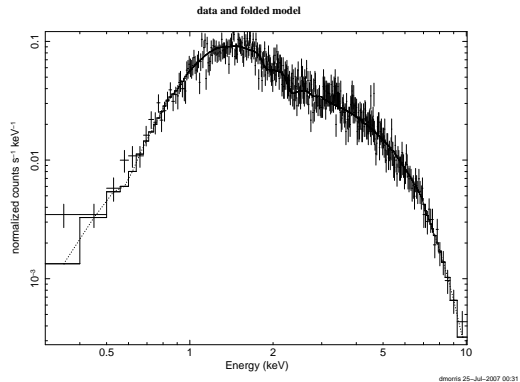}
  \end{center}
  \caption{The X-ray spectrum for Swift observations of LS I +61 303. The fit is for a pure power-law with photoelectric absorption.}
  \label{spectrum}
\end{figure*}

The source spectrum for the complete dataset is shown in
Figure~\ref{spectrum}. It is well fit by a simple power law plus
photoelectric absorption model ($\chi^2$/d.o.f.=377/370), with no
indication of spectral lines or a thermal black-body component. The
differential flux at 1~keV is
$1.334_{-0.020}^{+0.019}\times10^{-11}\mathrm{erg~cm^{-2}~s^{-1}}$, the photon
index is $1.73_{-0.03}^{+0.03}$ and the neutral Hydrogen column density is
$0.576_{-0.018}^{+0.019}\times10^{22}\mathrm{cm^{-2}}$. Further analysis will
examine spectral variations as a function of time and phase, and
explore the possibility of correlations between the spectral
parameters.

While there is no doubt that LS I +61 303 is a strongly variable X-ray
source, previous evidence for a consistent orbital modulation of the
X-ray flux is unconvincing, largely due to the sparseness of the
data. Our observations reaffirm that the system is a strongly variable
X-ray source. Preliminary analysis reveals no clear evidence for
variability in the hardness ratio or correlation between the hardness
ratio and flux.

Chernyakova et al. \cite{chernyakova06} have claimed evidence
for a systematic variation of the X-ray spectrum across the orbit,
with higher X-ray flux producing a harder spectrum, as well as for
significant variation in the column density, $N_H$. They explain this variation as
due to the erratic acceleration and cooling of electrons in the shocks
formed between the pulsar wind and an inhomogeneous ``clumpy wind''
from the Be star. Detailed, time resolved, spectral analysis of our
dataset will provide additional constraints to this, and other, models
for non-thermal emission from LS I +61 303.

\vspace{-0.4cm}
\section{Acknowledgements}
\vspace{-0.3cm}
This work is supported at Pennsylvania State University by NASA contract NAS 5-00136.
\vspace{-0.5cm}

\bibliography{icrc0233}
\vspace{-0.3cm}
\bibliographystyle{plain}

\end{document}